\documentstyle[twocolumn,aps,prd]{revtex}
\tighten

\def\bfl{\begin{flushleft}}
\def\efl{\end{flushleft}}
\def\bfr{\begin{flushright}}
\def\efr{\end{flushright}}
\def\bc{\begin{center}}
\def\ec{\end{center}}
\def\be{\begin{equation}}
\def\ee{\end{equation}}
\def\ba{\begin{eqnarray}}
\def\ea{\end{eqnarray}}
\def\baa#1{\begin{array}{#1}}
\def\eaa{\end{array}}
\def\nn{\nonumber }
\def\lb#1{\label{#1}}

\def\drm{d}

\def\ap{{\tilde A}}
\def\up{{\tilde U}}

\begin{document}

\wideabs{
\draft

\title{
~~~~~~~~~~~~~~~~~~~~~~~~~~~~~~~~~~~~~~~~~~~~~~~~~~~~~~~~~~~~~~
~~~~~~~~
{\footnotesize Phys.Lett. B527 (2002) 215-225 
~~
(hep-th/0102127)}\\
Core structure and exactly solvable models in 
dilaton gravity coupled to Maxwell and antisymmetric tensor fields
      }

\author{
Konstantin G. Zloshchastiev
}

\address{Department of Physics, National University of Singapore,
Singapore 117542, Republic of Singapore\\
and 
Department of Theoretical Physics, Dnepropetrovsk State University,
Dnepropetrovsk 49050, Ukraine
}

\date{~Received: ~~~~~~~~~~~~~~~~~~~~~~}
\maketitle

\begin{abstract}
We consider the D-dimensional massive dilaton gravity
coupled to Maxwell and antisymmetric tensor fields (EMATD).
We derive the full separability of this theory in static case.
This discloses the core structure of the theory and
yields the simple procedure of how to generate integrability classes.
As an example we take a certain new class, obtain the two-parametric
families of dyonic solutions.
It turns out that at some conditions they 
tend to
the D-dimensional dyonic Reissner-Nordstr\"om-deSitter solutions but with 
``renormalized'' dyonic charge plus a small logarithmic correction.
The latter
has the significant influence on the global structure of
the non-perturbed solution - it may shift and split horizons,
break down extremality, and dress the naked singularity. 
We speculate on physical importance of the deduced integrability
classes, in particular on their possible role in understanding of
the problem of unknown dilaton potential in 
modern cosmological and low-energy string models.

\end{abstract}


}


\narrowtext

\section{Introduction}\lb{s-in}

We start with the following action 
\be
S = 
\int \drm^D x \sqrt{- g} 
\biggl[
      R - \frac{\beta}{2} \,(\partial \phi)^2 
      + \Xi \,F^2 
      + \Psi \,F^2_{(p)} 
      + \Lambda 
\biggr],                                                   \lb{e-acti} 
\ee
where $R$ is the Ricci scalar,
$p=D-2$, $F$ and $F_{(p)}$ are two- and 
$p$-forms respectively, 
$\Xi$, $\Psi$  and $\Lambda$ are functions of dilaton $\phi$ \cite{lu93},
and $\beta$ is some unspecified constant (not necessary positive).
The models of such a kind appear in modern cosmological and
low-energy string and supergravity theories.
We will be interested in static solutions of this system
hence further we will work
with the metric ansatz
\be
\drm s^2 = - e^{U(r)} \drm t^2 + e^{-U(r)} \drm r^2 +
e^{A(r)} \drm s^2_{(p,k)},                         \lb{e-metr}
\ee
where  $\drm s^2_{(p,k)}$ is a $p$-dimensional maximally symmetrical space
with $k$ being $-1$, $0$, $+1$ depending on geometry 
(${\cal H}^p$, ${\cal E}^p$, ${\cal S}^p$) -
we are going to
handle them simultaneously and uniformly.
The Maxwell and $p$-form fields are assumed being in the form
\be
F = Q \frac{e^{-p A/2}}{\Xi}  \drm t \wedge \drm r, \ \ 
F_{(p) M...N} = P e^{-p A/2} \varepsilon_{M...N} 
\ee
in an orthonormal frame,
where $Q$ and $P$ are electric and magnetic charges.
Then the field equations become 
\ba
&&
A'' + A' (U' + p A'/2) - 
\frac{2}{p} \hat\Xi \, e^{-p A-U} 
- \frac{2}{p} \Lambda e^{-U} 
\nn\\&&\quad
- 2 k (p-1) e^{-A-U} =0
,                                                        \lb{e-em1}\\
&&
\beta  \phi'' + \beta \phi' (U'+ p A'/2) +  
\hat\Xi_{,\phi}\, e^{-p A-U} +
\Lambda_{,\phi}\, e^{-U} 
=0,                                                       \lb{e-em2}  \\
&&
A'' + A'^2/2 + \beta \phi'^2/p
=0,                                                        \lb{e-em3} 
\ea
where 
$' \equiv \partial_r$ and
$\hat\Xi \equiv 2 Q^2\,\Xi^{-1} + p! \, P^2 \, \Psi $.

The paper is arranged as follows.
In next section we prove the full separability of static
EMATD theory and reveal its core structure.
The latter is formulated in terms of the classes of integrability 
(more correctly, solvability).
Sec. \ref{s-lc} deals with a particular example of the so-called
linear class. 
We rule out the relevant models and their solutions' families,
and study the physical relevance of latter.
Sec. \ref{s-lt} is devoted to how the discovered core 
structure of EMATD gravity 
might help theorists in the situation when neither couplings
nor potential are known precisely.
Conclusions are made in Sec. \ref{s-co}.

\section{Separability and core structure of EMATD gravity}\lb{s-se}

Now we will rule out the full separability of the static
dyonic EMATD theory
(the case without antisymmetric tensor has been 
considered in ref. \cite{z0101}).
Due to that separability, we will be equipped 
with the straightforward procedure
of generating of the 
numerous classes of integrability that are dyonic besides
$\Lambda$ is non-zero in general case.

Applying the approach of ref. \cite{z0101} one can 
obtain the following system 
(which is similar to that from the EMD case)
\ba
&& 
2 k (p-1) + 
\frac{2 e^\ap}{p}
\left(
      \Lambda + e^{-p \ap} \hat\Xi
\right)
+
e^{\up + 2 Y}
\nn\\&&\quad
\times
\biggl(
      \frac{\beta}{p \ap_{,\phi}^{2}} 
-
\frac{\up_{,\phi}}{\ap_{,\phi}} 
- \frac{p-1}{2}
\biggr)
= 0, \lb{e-g1}\\
&& 
2 k (p-1) + 
\frac{2 e^\ap}{p}
\left(
      \Lambda + \frac{p }{2\beta} \Lambda_{,\phi} \ap_{,\phi}
\right)
+
e^{\up + 2 Y}
\biggl(\frac{1}{\ap_{,\phi}}
\biggr)_{,\phi}
\nn\\&&\quad 
+
\frac{2 e^{-(p-1) \ap}}{p}
\left(
      \hat\Xi + \frac{p }{2\beta} \hat\Xi_{,\phi} \ap_{,\phi}
\right)
= 0,                                               \lb{e-g2}\\
&&
\phi' \pm \frac{e^{Y-\ap/2}}{ \ap_{,\phi}} =0,  \quad
Y (\phi) \equiv - \frac{\beta}{p} 
\int \frac{\drm \phi}{ \ap_{,\phi}} + Y_0,       \lb{e-g3}
\ea
where $U (r) \equiv \up (\phi(r))$ and 
$A(r) \equiv \ap (\phi(r))$, 
the subscript ``$,\phi$'' stands for the derivative
with respect to dilaton.
This system is equivalent to the initial one but has much more 
capabilities.
First of all, it is explicitly separable:
$\up$ is algebraically given by Eq. (\ref{e-g2}) so one can easily
exclude it from Eq. (\ref{e-g1}) to receive the so-called
class equation
\ba                                            
&&
\frac{H_{,\phi}}{\ap_{,\phi}}  
+
\left(
\frac{\beta}{p \ap_{,\phi}^2} + \frac{p-1}{2}
\right)
H
+
k (p-1) 
\nn\\&& \qquad \qquad\qquad\qquad\quad
+ 
\frac{e^\ap}{p}
\left(
      \Lambda + e^{-p \ap} \hat\Xi
\right)=0,                         \lb{e-g1a}
\ea
where
$
H \equiv
\frac{1}{p (1/\ap_{,\phi})_{,\phi}}
\biggl[
k p (p-1) + 
e^\ap
\left(
      \Lambda + \frac{p }{2\beta} \Lambda_{,\phi} \ap_{,\phi}
\right)
+
e^{-(p-1) \ap}
\left(
      \hat\Xi + \frac{p }{2\beta} \hat\Xi_{,\phi} \ap_{,\phi}
\right)
\biggr]
,
$
and thus to
come to the system of autonomous equations
yielding $A$, $\phi$, $U$ consecutively.
Eq. (\ref{e-g1a}) is a non-linear
third-order ODE with respect to $\ap (\phi)$ 
so the direct task (finding of $\ap$ at given 
$\Xi$, $\Psi$, $\Lambda$) is still hard to solve
without supplementary symmetries or assumptions.
However, using this equation one may study the inverse problem,
i.e., the obtaining of the $\Xi$-$\Psi$-$\Lambda$ triplets 
corresponding to a concrete fixed $\ap$.
Thus, with every $\ap$ it is associated the appropriate class of 
integrability determined by the equation above.
It will help that the 
equation is a linear (at most) second-order ODE with respect to
$\hat\Xi$ and $\Lambda$,
besides, having only one equation 
it is much easier to study the integrability classes
numerically, e.g., to clarify whether they always have stable solutions,
see ref. \cite{mps88} and
references on appropriate methods therein.

Moreover, there is an exceptional class in this construction.
If $\ap \sim \phi$  
then $\up$ immediately disappears in Eq. (\ref{e-g2}),
so the latter becomes a  linear first-order ODE
with respect to $\Lambda$ and $\Xi$.
The linear class is of interest both by itself and
in connection with supergravity models,
so below we will study it in more details.
Then, as an example, we will
pick some concrete EMATD model to
obtain its general-in-class solutions.


\section{An example: linear class}\lb{s-lc}

Let us impose
\be                                       \lb{e-la}
\ap = \frac{4 d_1}{p} \phi - \ln d_2,
\ee
with $d_i$ being arbitrary constants.
Then Eq. (\ref{e-g3}) yields
\ba                                       \lb{e-lph}
\phi = 
\cases{
\frac{4 p d_1}{8 d_1^2 + p\beta}
\ln{
\left[
\frac{\sqrt{d_2} (8 d_1^2 + p\beta) (r-r_0)}
         {16 d_1^2}
\right]
   }, & $d_1^2 + \frac{p\beta}{8} \not=0$, \cr
-i \sqrt{\frac{d_2 p}{2\beta}} (r-r_0), & 
$d_1 = \frac{i}{2} \sqrt{\frac{p\beta}{2}}$, \cr
}
\ea
and Eq. (\ref{e-g2}) becomes the equation of integrability class
\ba
&&
\frac{e^{\frac{4 d_1}{p} \phi}}{d_2}
\left( \Lambda  + 
     \frac{2 d_1}{\beta } \,\Lambda_{,\phi}
\right)
+
\frac{e^{\frac{4 (1 -p) d_1}{p} \phi}
     }{d_2^{1-p}}
\left(
      \hat\Xi + \frac{2 d_1}{\beta}\, \hat\Xi_{,\phi}
\right)  
\nn\\&&\qquad
+k p (p-1) =0,                                    \lb{e-lcl}
\ea
whereas $\up$ can be easily found from Eq. (\ref{e-g1}) which is the
linear first-order ODE with respect to $e^\up$.
It should be noted that
the extended Lambda-Maxwell duality (discussed in ref. \cite{z0101} at 
$\Psi = 0$)
appears to be broken at $D \not= 4$;
it is curious that electric-magnetic duality is also
broken if $D \not= 4$, therefore, 
$D=4$  turns out to be a magic number again.
Now it is time to take some concrete
narrow physical system and obtain its
solutions within the frameworks of the linear class.


\bc \bf
String-inspired model: solutions
\ec

We choose the  physically important model which
was first integrated 
(at $\Lambda = 0$)
by Gibbons and Maeda \cite{gm88}:
\be                                            \lb{e-gmxi}
\Xi = - e^{-\frac{4 g_2}{p} \phi}, \ \
\Psi = - \frac{2}{p!} e^{-\frac{4 g_p}{p} \phi}, \ \
\beta = 8/p,
\ee
where $g$'s are coupling constants.
When  $\Psi$  vanishes then
$g_2 = 1$ corresponds to field theory limit of superstring model,
$g_2 = \sqrt{1+p/n}$ corresponds to the toroidal $T^n$ reduction
of ($D+n$)-spacetime to $D$-spacetime,
$g_2 = 0$ is a usual Einstein-Maxwell system); 
$\Lambda$ is precisely unknown in string theory \cite{lu93}.
The models of such a type have been intensively studied in the
case $\Lambda = 0$ \cite{sev-massl} but the progress in the models which
contain both an antisymmetric tensor and a massive dilaton is still rather 
slow \cite{sev-massv} despite their obvious importance.
With the settings (\ref{e-gmxi}) Eq. (\ref{e-lph}) becomes
\ba                                       
\phi = 
\cases{
\frac{p d_1}{2 (d_1^2 + 1)}
\ln{
\left[
\frac{\sqrt{d_2} (d_1^2 + 1) (r-r_0)}
         {2 d_1^2}
\right]
   }, & $d_1^2 + 1 \not=0$, \cr
-i \frac{\sqrt{d_2}}{4} p (r-r_0), & 
$d_1 = i$ . \cr
}
                                     \lb{e-gmph}
\ea
For further it is convenient to define the three polynomials:
\[
\pi_1 = d_1^2 -1, \
\pi_2 = p d_1^2 - g_2 d_1 -1,  \
\pi_3 = p d_1^2 + g_p d_1 -1.
\]
The integration of the class equation above reveals the following
cases (note that additionally each case may contain the multiple 
subcases determined by the combinations of parameters apart
from $\pi_i =0$ ones at which an initial $\up$,
but not $\Lambda$, becomes singular):

(i) $\pi_1 \not=0$, $\pi_2 \not=0$, $\pi_3 \not=0$.\\
Integrating Eq. (\ref{e-lcl}) with (\ref{e-gmxi}) we see that $\Lambda$
must be
\ba
&&\Lambda = a_0 
e^{-\frac{4}{p d_1} \phi}
+ \frac{k d_2 }{\pi_1} p (p-1)
e^{-\frac{4 d_1 }{p}\phi}
- 2 d_2^p 
e^{-4 d_1 \phi}
\nn\\&&\qquad
\times
\biggl[
\frac{Q^2}{\pi_2} (1+ g_2 d_1) e^{\frac{4 g_2 }{p}\phi}
+
\frac{P^2}{\pi_3} (1 - g_p d_1) e^{-\frac{4 g_p}{p}\phi}
\biggr],                                                \lb{e-ilam}
\ea
where $a_0$ is integration constant,
corresponding $A,\, \phi$ are given by 
Eqs. (\ref{e-la}), (\ref{e-gmph}),
and $U$ is given by Eq. (\ref{e-g1}) provided (\ref{e-la})
and (\ref{e-gmxi}):
\ba
&&e^\up =
c e^{2\frac{1-(p-1) d_1^2}{p d_1}\phi}
+
\frac{4 k (p-1) d_1^4 e^{\frac{4\phi}{p d_1}}}
     {\pi_1 (p d_1^2 - \pi_1)}
+
\frac{4 a_0 d_1^2 e^{\frac{4 d_1}{p}\phi}}
     {d_2 p (p d_1^2 + \pi_1)}
\nn\\&&\quad
+
\frac{8 Q^2 d_1^4 d_2^{p-1}  e^{-\frac{4(\pi_2 - d_1^2)}{p d_1}\phi}}
     { \pi_2 (\pi_2 - d_1(d_1 + g_2))}
+
\frac{8 P^2 d_1^4 d_2^{p-1}  e^{-\frac{4(\pi_3 - d_1^2)}{p d_1}\phi}}
     { \pi_3 (\pi_3 - d_1(d_1 - g_p))},
\ea
where $c$ is another integration constant related to mass.
As was alerted above, this case contains subcases 
$p d_1^2 \pm \pi_1=0$,  $\pi_2 - d_1(d_1 + g_2)=0$,
$\pi_3 - d_1(d_1 - g_p) = 0$ making the last equation, but not 
Eq. (\ref{e-ilam}), singular.
For the sake of brevity, we do not present them here.

(ii) $\pi_1 =0$, $\pi_2 \not=0$, $\pi_3 \not=0$.\\
We choose the positive root $d_1 = 1$ then
in the same way as above one can show that $\Lambda$
must be
\ba
&&\Lambda = 
\left[
a_0 
- 4 k d_2 (p-1) \phi
\right]
e^{-\frac{4}{p} \phi}
- 2 d_2^p 
e^{-4 \phi}
\nn\\&&\qquad
\times
\biggl[
\frac{Q^2 (1+ g_2) e^{\frac{4 g_2 }{p}\phi}}
     {p-1-g_2} 
+
\frac{P^2 (1 - g_p) e^{-\frac{4 g_p}{p}\phi}}
     {p-1+g_p} 
\biggr],
\ea
corresponding $A,\, \phi$ are given by 
Eqs. (\ref{e-la}), (\ref{e-gmph}) with $d_1$ being as above,
whereas $U$ turns out to be
\ba
&&e^\up =
\frac{e^{\frac{4}{p}\phi}  }{(p/2)^2}
\left[
c e^{-2\phi}
+ \frac{a_0}{d_2}
+ k (p^2+p-2) - 4 k (p-1) \phi
\right]
\nn\\&&\
+
\frac{8 Q^2 d_2^{p-1}  e^{-\frac{4(p-2-g_2)}{p}\phi}}
     {(p-2-2 g_2)(p-1-g_2)}
+
\frac{8 P^2 d_2^{p-1}  e^{-\frac{4(p-2+g_p)}{p}\phi}}
     {(p-2+2 g_p)(p-1+g_p)}
.                \nn
\ea

(iii) $\pi_1 \not=0$, $\pi_2 =0$, $\pi_3 \not=0$.\\
To avoid root branches we will work in terms of $d_1$
assuming that it is related to $g_2$ via the
relation $g_2 = p d_1 - 1/d_1$,
then $\Lambda$ is
\ba
&&\Lambda = 
\left(
a_0 
+ 8 Q^2 d_1 d_2^p \phi
\right)
e^{-\frac{4}{p d_1} \phi}
+
\frac{k d_2 p }{\pi_1} (p-1)
e^{-\frac{4 d_1}{p} \phi}
\nn\\&&\qquad
-
\frac{2 P^2 d_2^p }{\pi_3}
(1-g_p d_1)
e^{-\frac{4 (p d_1 + g_p)}{p} \phi}
,
\ea
corresponding $A,\, \phi$ are given by 
Eqs. (\ref{e-la}), (\ref{e-gmph}) with $d_1$ being as above,
whereas $U$ turns out to be
\ba
&&e^\up =
c e^{2\frac{1-(p-1)d_1^2}{p d_1}\phi}
+ 
\frac{4 k d_1^4 (p-1) e^{\frac{4\phi}{p d_1}}}
     {\pi_1 (pd_1^2 - \pi_1)}
+
\frac{8 Q^2 d_1^2 d_2^{p-1}}
     {p d_1^2 +\pi_1}
\nn\\&&\quad
\times
\frac{e^{\frac{4 d_1}{p} \phi}  }
     {p}
\left[
4 d_1 \phi
+ \frac{a_0}{2 Q^2 d_2^p}
- \frac{3 p d_1^2 +\pi_1}{p d_1^2 +\pi_1}
\right]
\nn\\&&\quad
+
\frac{
      8 P^2 d_1^4 d_2^{p-1}
      e^{-4\frac{\pi_3 - d_1^2}{p d_1}\phi}
     }
     {\pi_3 (\pi_3 - d_1( d_1 - g_p))}
.                \lb{e-iiiu}
\ea

(iv) $\pi_1 \not=0$, $\pi_2 \not=0$, $\pi_3 =0$.\\
This case is identical to the previous one if one replaces everywhere
$g_2$ with $-g_p$ and interchanges $Q$ and $P$.

(v) $\pi_1 =0$, $\pi_2 =0$, $\pi_3 \not=0$.\\
Therefore, we have the following two sets
$\{g_2,\, d_1\} = \pm \{ p-1,\, 1 \}$.
We choose the plus branch
then $\Lambda$ is
\ba
&&\Lambda = 
\left[
a_0 
+ 4 d_2 (2 Q^2 d_2^{p-1} - k (p-1)) \phi
\right]
e^{-\frac{4}{p} \phi}
\nn\\&&\qquad
-
\frac{2 P^2 d_2^p (1- g_p)}{p-1+g_p}
e^{-\frac{4 (p + g_p)}{p} \phi}
,
\ea
corresponding $A,\, \phi$ are given by 
Eqs. (\ref{e-la}), (\ref{e-gmph}) with $d_1$ being as above,
whereas $U$ turns out to be
\ba
&&e^\up =
\frac{e^{\frac{4}{p}\phi}  }
     { p^2}
\Biggl[
c e^{-2 \phi}
+
4 
\left(
\frac{a_0}{d_2} 
+ k (p^2+p-2) - \frac{6 Q^2}{d_2^{1-p}}
\right)
\nn\\&&
+
16 \phi 
\left(
\frac{2  Q^2}{d_2^{1-p}} - k (p-1)
\right)  
\Biggr]
+
\frac{
      8 P^2 d_2^{p-1}
      e^{4\frac{2-p-g_p}{p}\phi}
     }
     {(p-1+g_p) (p-2 + 2 g_p)}
.                \nn
\ea

(vi) $\pi_1 =0$, $\pi_2 \not=0$, $\pi_3 =0$.\\
Similarly, we have the following two sets
$\{g_p,\, d_1\} = \pm \{ 1-p,\, 1 \}$.
One can use the expressions from the previous case but 
has to replace in them $g_p$ with $-g_2$ and interchange $Q$ and $P$.

(vii) $\pi_1 \not=0$, $\pi_2 =0$, $\pi_3 =0$.\\
It contains the condition $g_2 + g_p =0$,
so we can exclude $g_p$. 
Besides, to avoid root branches we will work again in terms of $d_1$
assuming that it is related to $g_2$ via the
relation $g_2 = p d_1 - 1/d_1$.
We have
\ba
&&\Lambda = 
\left[
a_0 
+ 8 d_1 d_2^p (Q^2+P^2) \phi
\right]
e^{-\frac{4 \phi}{p d_1} }
+
\frac{k d_2 p (p-1)}
     {\pi_1
e^{\frac{4 d_1}{p } \phi}   }
,
\ea
corresponding $A,\, \phi$ are given by 
Eqs. (\ref{e-la}), (\ref{e-gmph}) with $d_1$ being as above,
whereas for $U$ one can formally use Eq. (\ref{e-iiiu}) without the last 
term ($\sim P^2$) and with $Q^2$ being replaced
with $Q^2+P^2$.

(viii) $\pi_1 =0$, $\pi_2 =0$, $\pi_3 =0$.\\
Therefore, $g_2+g_p=0$ and we have the following two sets
$\{g_2,\, g_p,\, d_1\} = \pm \{ p-1, \, 1-p, \, 1 \}$.
We choose the plus branch then $\Lambda$ is
\ba
&&\Lambda = 
\biggl[
a_0 
+ 4 d_2 
\left(
2 d_2^{p-1} (Q^2+P^2)  - k (p-1)
\right) \phi
\biggr]
e^{-\frac{4}{p} \phi}
,
\ea
corresponding $A,\, \phi$ are given by 
Eqs. (\ref{e-la}), (\ref{e-gmph}) with $d_1$ being as above,
whereas for $U$ one can formally use the corresponding
expression from (v) but without the last 
term ($\sim P^2$) and with $Q^2$ being replaced
with $Q^2+P^2$.

Thus, we have enumerated the basic solutions, which correspond
to the model (\ref{e-gmxi}) within frameworks of the linear class.
Of course, we have mentioned just a few examples.


\bc \bf
String-inspired model: discussion of solutions
\ec

Analyzing the solutions above, one can see that
the set (i) 
is the largest set of solutions
due to the parameter $d_1$ being non-fixed there.
In this section we will study the solutions (i) in details.
In view of future considerations, let us first redefine the constants
\be
c = - 8\mu d_2^{\frac{p_1 + d_1^{-2}}{2}}, \ \
a_0 = \Lambda_0 d_2^{d_1^{-2}},
\ee
where $p_1 \equiv p-1$.
The next step is to switch coordinates to the
infinite-observer frame of reference
\be
e^{A(r)} = 
\left[
\frac{1+d_1^{2}}{2 d_1^2 d_2^{\frac{1}{2} d_1^{-2}}} r
\right]^{\frac{2 d_1^2}{1+d_1^2}}
\to r^2, \ \
2 d_2^{\frac{1}{2} d_1^{-2}} t \to  t,
\ee
then in new coordinates we obtain 
\ba
&&\drm s^2 = - 
\drm t^2 
\biggl[
\frac{k p_1 d_1^4  r^{2 d_1^{-2}}}
     {\pi_1 (p_1 d_1^2 +1)} 
-
\frac{2 \mu}{ r^{p_1-d_1^{-2}}}
+
\frac{\Lambda_0 d_1^2 r^2}{p((p+1) d_1^2 - 1)}
\nn\\&& \ \
+ \frac{2 d_1^4 d_2^{g_2 d_1^{-1}}
        Q^2 r^{2(1 - \pi_2 d_1^{-2})}}
       {\pi_2 (p_1 d_1^2 - 2 g_2 d_1 - 1)}
+ \frac{2 d_1^4 d_2^{- g_p d_1^{-1}}
        P^2 r^{2(1 - \pi_3 d_1^{-2})}}
       {\pi_3 (p_1 d_1^2 + 2 g_p d_1 - 1)}
\biggr] 
\nn\\&& \
+ 
\drm r^2 
\biggl[
\frac{k p_1 d_1^4}
     {\pi_1 (p_1 d_1^2 +1)} 
-
\frac{2 \mu}{ r^{p_1 + d_1^{-2}}}
+
\frac{\Lambda_0 d_1^2 r^{2 \pi_1 d_1^{-2}}}{p((p+1) d_1^2 - 1)}
\nn\\&& \ 
+ \frac{2 d_1^4 d_2^{g_2 d_1^{-1}}
        Q^2 r^{2(g_2 d_1^{-1}-p_1)}}
       {\pi_2 (p_1 d_1^2 - 2 g_2 d_1 - 1)}
+ \frac{2 d_1^4 d_2^{- g_p d_1^{-1}}
        P^2 r^{2(p_1 + g_p d_1^{-1})}}
       {\pi_3 (p_1 d_1^2 + 2 g_p d_1 - 1)}
\biggr] 
\nn\\&& \ 
+ r^2 \drm s^2_{(p,k)},  
\quad
e^\phi = 
\left(
\sqrt{d_2} r
\right)^{\frac{p}{2} d_1^{-1}}, 
\nn\\&& 
F = 
\frac{ d_2^{g_2 d_1^{-1} } Q}
     {r^{p-2 g_2 d_1^{-1} - d_1^{-2}}}  \drm r \wedge \drm t, \ \ 
F_{(p) M...N} = \frac{P}{r^p} \varepsilon_{M...N}.      \lb{e-ifld}
\ea
The remainder of this section will be devoted to the studies 
of this solution at non-fixed large values of $|d_1|$.
Assuming $|d_1| \gg \text{max} \{1,\, |g_2|,\, |g_p| \}$, we obtain
that up to the order $O[1/d_1^{2}]$ 
(here and below
it is supposed to be the default precision of calculations)
the metric above takes the 
habitual form
\[
\drm s^2 = - e^{U(r)} \drm t^2 + e^{-U(r)} \drm r^2 +
r^2 \drm s^2_{(p,k)},                         
\]
with
\be                                                \lb{e-ieuapp}
e^{U(r)} =
k - \frac{2 \mu}{r^{p_1}} +
\frac{\Lambda_0 r^2}{p(p+1)}
+ \frac{\Delta - \Theta \ln{(r^{p_1}/\eta)}}{r^{2 p_1}},
\ee
where we have defined the following constants
\ba
&&
\Delta = \frac{2}{p p_1}
\left(
Z^2
+
\frac{3 p -1}{p \, p_1 \, d_1} W
\right), \ \
\Theta = -
\frac{4 W}{p\, p_1^2 d_1},
\nn\\&& 
\eta = 
d_2^{-\frac{p_1}{2}},  \ \
W =  g_2 Q^2 - g_p P^2, \ \
Z^2 = Q^2 + P^2,  
\ea
and it is implied that $p > 1$ 
(lower-dimensional cases will be separately considered after).
Also, the $O[d_1^{-2}]$-asymptotical form of the dilaton 
potential (\ref{e-ilam}) 
is
\be
\Lambda = \Lambda_0 - \frac{2 d_2^p W}{p d_1}
e^{-4 d_1 \phi}
.
\ee
The first, second and third terms in the metric above is
the D-dimensional Schwarzschild-deSitter.
The term proportional to $\Delta$ is nothing but the D-dimensional
Reissner-Nordstr\"om with the only difference
that the effective dyonic charge 
is the standard one plus a small correction of order $d_1^{-1}$.
The last term, proportional $\Theta$, is definitely something new,
and below we will study its influence in details.

From now we will work with the spherical case $k=1$, besides
we will neglect the cosmological constant for simplicity.
Then the information about 
the global structure of the metric can be read off
from the intersection of two curves described by the
following algebraic equation
\be                                        \lb{e-ihore}
x^2 - 2 \mu x + \Delta
\equiv
(x - \delta_+ ) (x - \delta_-) 
= \Theta \, \ln{(x/\eta)},
\ee
where $x = r^{p_1}$ and $\delta_\pm = \mu (1 \pm \sqrt{1 - \Delta/\mu^2})$.
It is useful to keep in mind that $\Theta$ is small ($\sim d_1^{-1}$)
that simplifies subject matter.
This smallness in fact means that 
for the whole
region except perhaps $x \to 0$ and $x \to + \infty$
the value of the logarithm in the equation
above should be assumed small in comparison with the
parameters $\mu$, $\Delta$ and $\eta$.

Case  $\mu^2 > \Delta$.
If $\Theta \equiv 0$ (that may happen not only when $d_1 =\infty$
but also when $g_2 Q^2 = g_p P^2$) 
this case corresponds to the D-dimensional
Reissner-Nordstr\"om  black hole.
Otherwise we have to solve the transcendental equation (\ref{e-ihore})
with real $\delta$'s.
Fortunately, it can be done analytically 
with the use of $\Theta$'s smallness.
Solving it, we obtain that we still have two horizons but their
radii acquire a correction:
\be
r_{H^\pm} = 
\left[
\delta_\pm + \frac{\Theta \ln{(\delta_\pm /\eta)}}
                 {2 (\delta_\pm - \mu)}
\right]^{\frac{1}{p_1}},
\ee
and the corresponding Hawking temperatures are calculated to be
\ba
&&
T_{H^\pm} = 
\frac{p_1 \delta^{-\frac{p}{p_1}}}{2 \pi} 
\biggl[
\delta_\pm - \mu
-
\frac{\Theta}{2 \delta_\pm}
\nn\\&& \ \quad           \times
\left(
1+
\frac{
      (\delta_\pm - p \mu)
      \ln{(\delta_\pm/\eta)}
     }
     {p_1 (\delta_\pm-\mu)}
\right)
\biggr]
,
\ea
an absolute value is implied.

Case  $\mu^2 = \Delta$.
Without the $\Theta$-perturbation 
this case corresponds to the D-dimensional
extremal Reissner-Nordstr\"om  black hole.
It turns out that the series expansion used in the previous case
fails (diverges) so we have to invent another one.
The non-perturbed horizon appears at $x=\mu$.
We are interested in small deviations from the non-perturbed case
so it is natural to expand Eq.(\ref{e-ihore}) with respect
to $x$ up to the
third order near this point.
We obtain that Eq.(\ref{e-ihore}) becomes
the quadratic equation,
\[                                     
\left(
1+ \frac{\Theta}{2 \mu^2}
\right) x^2
- 2 \mu
\left(
1+ \frac{\Theta}{\mu^2}
\right) x
+ \mu^2
= \Theta
\left[
\ln{(\mu/\eta)}
- 
\frac{3}{2} 
\right] ,
\]
from which one concludes that extremality is broken and the
extreme horizon is shifted and split into two ones, with the radii
\be
r_{H^\pm} = 
\left[
\mu + \frac{\Theta}{2 \mu}
\pm \sqrt{\Theta \ln{(\mu/\eta)}}
\right]^{\frac{1}{p_1}}.
\ee
Here, the term proportional to $\Theta$ shifts the horizon outward
or inward (depending on the sign of $W/d_1$)
whereas the term proportional to $\sqrt\Theta$ describes the split.
It is curious that in the particular case $\eta = \mu$ the extremality
is again restored up to $O[d_1^{-2}]$.
The corresponding Hawking temperatures are 
\be
T_{H^\pm} = 
\frac{
      \sqrt{\Theta \,\ln{(\mu/\eta)}      }
     }
     {2\,\pi 
      \mu^{\frac{p+p_1}{p_1}} 
     }
\left[ 
   p_1 \,\mu  \pm 
   p  \sqrt{\Theta \ln{(\mu/\eta)} }
\right]
,
\ee
and they do vanish not only when $d_1 = \infty$ but also at $\eta= \mu$.

Case $\mu^2 < \Delta$.
If $\Theta \equiv 0$ then the solution describes the  naked
Reissner-Nordstr\"om singularity.
There is a strong hope that the $\Theta$-perturbation
``dresses'' the singularity, i.e., creates a horizon around it.
To prove it, one has to show the conditions at which
the parabola and logarithmic curve have the intersection point(s)
even if the former does not cross an $x$-axis.
The intuitive solution for this is to require the minimum point of the
parabola to be as closely as possible to the $x$-axis, 
hence, to the logarithmic curve, because the latter is small.
The distance from the minimum point of the
parabola to the $x$-axis equals to $\Delta - \mu^2$,
so $\Delta$ must be equal to $\mu^2$ plus a small positive
correction, say
\be                                         \lb{e-idelta3}
\Delta = \mu^2 + |\text{const} \, d_1^{-1}|.
\ee
Again, we expand Eq. (\ref{e-ihore}) near the minimum point
of parabola and obtain the quadratic equation
\[                                    
\left(
1+ \frac{\Theta}{2 \mu^2}
\right) x^2
- 2 \mu
\left(
1+ \frac{\Theta}{\mu^2}
\right) x
+ \Delta
= \Theta
\left[
\ln{(\mu/\eta)}
- 
\frac{3}{2} 
\right] .
\]
If it has complex roots then the singularity is naked
otherwise it is hidden under at least one horizon.
One can check that this equation in general case does not
have real roots but if $\Delta$ is
\[
\Delta =
{\mu }^2 - \frac{2\,\Theta \,{\mu }^2\,\ln (\mu/\eta)}
   {\Theta  - 4\,{\mu }^2}
= \mu^2 + \frac{\Theta}{2} \ln{(\mu/\eta)},
\]
i.e., of the form (\ref{e-idelta3}),
provided $d_1^{-1} W \ln{(\mu/\eta)}$ is non-positive,
then the imaginary part vanishes, 
so one does have the purely real double root.
It means that we have found an example when a singularity is
dressed by the single horizon.
Its radius is
\be
r_{H^\pm} = 
\left[
\mu + \frac{\Theta}{2 \mu}
\right]^{\frac{1}{p_1}},
\ee
but with the Hawking temperature,
\be                                             \lb{e-ith3}
T_{H} = 
\frac{
      p_1 \Theta \,\ln{(\mu/\eta)}    
     }
     {4\,\pi 
      \mu^{\frac{p+p_1}{p_1}} 
     }
,
\ee
being of order $O[d_1^{-1}]$, rather than $O[d_1^{-1/2}]$
as in previous case.

As a final part of this section, we have to study the low-dimensional
case.
Indeed, the majority of 
Eqs. (\ref{e-ieuapp})-(\ref{e-ith3}) are not applicable
when $D=3$ or $2$, i.e., when the number of spatial dimensions
is, respectively, two and one.
The two-dimensional case is of no interest here because all the
solutions were derived assuming $p \not= 0$ for obvious reasons.
In the 3D case when $|d_1|$ is large, instead
of Eq. (\ref{e-ieuapp}) we obtain
\be                                                \lb{e-3deu}
e^{U(r)} =
\zeta  - 2\,Z^2\,\ln (r \,{\sqrt{{d_2}}}) + 
  \frac{\Lambda_0 r^2}{2},
\ee
where $Z^2$ is as above and it is denoted
\[
\zeta =
\frac{g_p^2 Q^2 + g_2^2 P^2}{2 g_2^2 g_p^2}
-\frac{d_1 (g_p Q^2 - g_2 P^2)}{g_2 g_p}
- Z^2
-2 \mu,
\]
$d_2$ is assumed positive for 
definiteness.
The scalar, Maxwell and $p$-form fields (\ref{e-ifld}) 
do not undergo principal
changes in the sense that they are not singular when $p\to 1$.
However, it is worth to note that $p$-form becomes the plain
vertex-type vector field with the only non-zero component
$F_{(1)} = f(r)\, \drm \varphi$ where 
$\varphi$ is an angular coordinate.
It cannot be represented as an external derivative of
some potential, therefore, it is not possible to derive it
from the 3D variational principle - in the action (\ref{e-acti}) it may
appear only as a non-dynamical source term
\be 
\Psi \, F_{(1)\, \alpha} F_{(1)}^{~\alpha}
= \Psi \, f(r)^2 = \widehat\Lambda (\phi),
\ee
because dilaton is an invertible function of $r$.
Nevertheless, one can consider the 1-form contribution formally, 
so below we will not impose $P \equiv 0$.

The solution is essentially cosmological -
the metric (\ref{e-3deu}) tends to de Sitter one
(provided $\zeta$ is positive), and has the only singularity at $r=0$
provided $Z \not= 0$.
Its global structure is, however, non-trivial and
crucially depends on values of the parameters.
Simple analysis shows that:
(a) when $\Lambda_0$ is negative, we have a single horizon
regardless of what other parameters are;
(b) when $\Lambda_0$ is positive, we have the naked singularity,
one extreme horizon, two horizons, depending on
whether the value
\[
\zeta + Z^2 - Z^2 \ln{(2 d_2 Z^2/ \Lambda_0)}
\]
is  positive, zero or negative, respectively;
(c) when $\Lambda_0$ vanishes, we have 
single horizon with 
\be
r_H = \sqrt{d_2\, e^{\zeta/ Z^{2}}  },
\ee
but the solution is not asymptotically flat 
(despite the curvature invariants do vanish asymptotically) so 
this is still cosmological, rather than black hole, horizon.

To summarize this section:
we have demonstrated that the two-parametric family of exact solutions (i)
at large values of one of the parameters describes the solution
which is
the D-dimensional Reissner-Nordstr\"om-deSitter solution but with 
``renormalized'' dyonic charge plus a perturbative non-constant
(logarithmic) correction  (\ref{e-ieuapp}).
It is also shown that this correction despite its smallness
has significant influence on the global structure of
the non-perturbed solution - it may shift and split horizons,
break down extremality, and dress the naked singularity.

\section{Integrability 
classes and dilaton potential problem}\lb{s-lt}

In the previous section we studied some particular class
as the fruitful example of the proposed approach's power.
Other integrability classes (not talking about models) 
are so diverse and numerous that in principle never can be 
covered all.
Other $\Xi$, $\Psi$, $\Lambda$ that may appear from a concrete
problem can be  paired up within our class in a similar manner.
Despite this pairing is an artificial procedure the
generated exact 
solutions are better than numerical studies from
scratch, besides ones can verify or falsify qualitative
approaches and results.
Be that as it may, 
the good news is that now one has a powerful tool to study
numerous models in a straightforward and uniform way,
even having no background
in the theory of differential equations. 

Now, it is a good time to coin the advantages that come
after the proven separability of static EMATD theory.
This section will be devoted to the generic physical significance of the
integrability classes given by relations between $A$ and dilaton
alike  (\ref{e-g1a}).
Here we are going to justify  the point that 
the integrability classes of such a kind
is not only a mathematical object
but also can play key role in some fundamental aspects of
field theory and theoretical cosmology.

For instance, the integrability classes may help with
the problem of unknown dilaton potential $\Lambda$
(but, of course, by themselves they cannot provide a complete answer).
When supersymmetry is unbroken then the dilaton is in the same
supermultiplet as the graviton and hence cannot acquire a mass.
However, in low-energy region the supersymmetry is broken so 
the assumption $\Lambda \equiv 0$ is inconsistent with 
observations \cite{will}.
Nowadays it is the strong problem that the dependence of $\Lambda$ 
on scalar field is precisely known
neither from string theory nor from cosmology-related experiments.
The one of the ideas of getting $\Lambda (\phi)$ comes from
supersymmetry and supergravity - despite 
the theories with the dilaton potential of general type
are non-supersymmetric as a rule, some non-trivial potentials
can be justified by supergravity models.
The flaw, however, is that supergravity cannot predict dilaton potential
uniquely - it has been already proposed an enormous amount of 
them \cite{susypot}.

The integrability classes, which are incidentally based on dependence
$A(\phi)$, bring the view on the $\Lambda$-problem from the
viewpoint different from the above-mentioned ones.
First, one should make the important observation 
that the dependence $A(\phi)$ is 
more universal
than, e.g., $U(\phi)$ or $A(U)$.
Indeed, 
if the metric is in the gauge (\ref{e-metr}) then
$A \sim \ln{\text{det} g}$ is related to the radius of (compact)
factor space and thus  
determines the geometrical scale 
of extra $p$ dimensions.
On the other hand, the dilaton field  was introduced
historically to consider the gravitational constant as a variable,
and it describes thus the rigidity of spacetime.
Therefore, $A(\phi)$ symbolizes the dependence
\be
A(\phi) \sim \text{graviton-scale} (\text{dilaton-scale}),
\ee
or, the ``size(rigidity)'' one.
In view of this, the 
existence proof of above-mentioned integrability classes
claims that {\it with each such a dependence it is associated
a unique $\Xi$-$\Psi$-$\Lambda$ triplet}. 
Let us for clarity disregard the $p$-form, as in ref. \cite{z0101}.
Then, if one knows both the dependence $A(\phi)$ and the explicit form of
the dilaton-Maxwell coupling $\Xi$ then $\Lambda$
is uniquely determined by the class equation.
Occasionally, in our case 
$\Xi$ is known from (perturbative) string theory
so the problem now is what is the explicit relation of $A$ to dilaton.

So far, we do not have any clear idea on the latter.
But we sure that the above-studied linear class 
(and, therefore, models therein)
is at least
a first-order approximation 
if one expands the yet unknown
for sure True Function $A(\phi)$
in Taylor series with respect to dilaton.
It should be also noted that 
the linear class is distinct from others not only
because it is given by a first- rather than second-order ODE
with respect to $\Xi$ and $\Lambda$  but also because it possesses
a certain discrete symmetry that 
alike the electric-magnetic duality
in pure EMD is broken at $D \not= 4$ and thus it forces $D=4$ to be
a magic number again, see the paragraph after
Eq. (\ref{e-lcl}) and ref. \cite{z0101}.

\section{Conclusion}\lb{s-co}

We have deduced the full separability of the static
D-dimensional massive dilaton gravity
coupled to Maxwell and antisymmetric tensor fields for
arbitrary dilaton potential and dilaton-Maxwell and
dilaton-tensor coupling.
This fact allowed us to achieve the two following aims.

First, in Sec. \ref{s-se} 
the core structure of the theory has been revealed.
It turned out that it is the universal relation between 
characteristic scales of gravity and scalar field that lies in the
very heart of EMATD gravity and, probably, of any other 
Einstein gravity that contains scalar field.
In Sec. \ref{s-lt} we have demonstrated how the knowledge of this
structure can help us with the situation when neither coupling
nor potential are known precisely.
Then the core structure of EMATD gravity
suggests the self-consistency requirement:
with each above-mentioned relation it is associated
a unique dilaton coupling-coupling-potential ``triplet''. 
Therefore, if one knows couplings then one can determine potentials
and vice versa.

Second, separability has also led us to the practical
concept of integrability classes
that appeared to be powerful tool for getting of exactly solvable
EMATD models and their solutions.
As an example, we have studied the new class of solutions in
Sec. \ref{s-lc}.
It has been observed that in 4D this class obeys a certain
duality between dilaton potential and dilaton-Maxwell coupling.
There we have studied 
the physical properties of the two-parametric
family of dyonic solutions for the case of 
the exponential Gibbons-Maeda couplings
related to higher-dimensional gravities including superstrings.
It turned out that these solutions 
at some conditions resemble
the D-dimensional dyonic Reissner-Nordstr\"om-deSitter solutions but with 
``renormalized'' dyonic charge plus a small logarithmic correction.
The latter
has the significant influence on the global structure of
the non-perturbed solution - it may shift and split horizons,
break down extremality, and dress the naked singularity.

\section*{Acknowledgments}
I am grateful to Edward Teo 
(DAMTP, Univ. of Cambridge and Natl. Univ. of Singapore)
and Cristian Stelea (Univ. ``Alexandru Ioan Cuza'', Iasi, Romania 
and Nat'l. Univ. of Singapore)
for suggesting the theme and helpful discussions.

\def\AnP{Ann. Phys.}
\def\APP{Acta Phys. Polon.}
\def\CJP{Czech. J. Phys.}
\def\CMPh{Commun. Math. Phys.}
\def\CQG {Class. Quantum Grav.}
\def\EPL  {Europhys. Lett.}
\def\IJMP  {Int. J. Mod. Phys.}
\def\JMP{J. Math. Phys.}
\def\JPh{J. Phys.}
\def\FP{Fortschr. Phys.}
\def\GRG {Gen. Relativ. Gravit.}
\def\GC {Gravit. Cosmol.}
\def\LMPh {Lett. Math. Phys.}
\def\MPL  {Mod. Phys. Lett.}
\def\NPh  {Nucl. Phys.}
\def\PhE  {Phys.Essays}
\def\PhL  {Phys. Lett.}
\def\PhR  {Phys. Rev.}
\def\PhRL {Phys. Rev. Lett.}
\def\PhRp {Phys. Rept.}
\def\NCim {Nuovo Cimento}
\def\TMF {Teor. Mat. Fiz.}
\def\prp {report}
\def\Prp {Report}

\def\jn#1#2#3#4#5{{#1}{#2} {#3} {(#5)} {#4}}   

\def\boo#1#2#3#4#5{ #1 ({#2}, {#3}, {#4}){#5}}  



\begin{references}


\bibitem{lu93}
M. Cveti\v c and A. A. Tseytlin,
\jn{\NPh}{ B}{416}{137}{1994};
K.-L. Chan, 
\jn{\MPL}{ A}{12}{1597}{1997};
J.~X.~Lu,
\jn{\NPh}{ B}{409}{290}{1993};
J.H. Horne and G.T. Horowitz,
\jn{\NPh}{ B}{399}{169}{1993};
R. Gregory and J. A. Harvey,
\jn{\PhR}{ D}{47}{2411}{1993}.

\bibitem{z0101}
K. G. Zloshchastiev, 
\jn{\PhR}{ D}{64}{084026}{2001}.

\bibitem{mps88}
S.G. Matinyan, E. B. Prokhorenko, and G.K. Savvidy, 
\jn{\NPh}{ B}{298}{414}{1988}.


\bibitem{gm88}
G. W. Gibbons and K. Maeda, 
\jn{\NPh}{ B}{298}{741}{1988}.


\bibitem{sev-massl}
B.~Zhou and C.~Zhu,
\jn{Commun. Theor. Phys.}{}{32}{173}{1999};
K. S. Stelle, hep-th/9803116;
T.~Maki and K.~Shiraishi,
\jn{\CQG}{}{11}{2781}{1994};
M.~J.~Duff and J.~X.~Lu,
\jn{\NPh}{ B}{416}{301}{1994};
G.~T.~Horowitz and A.~Strominger,
\jn{\NPh}{ B}{360}{197}{1991};
A.~Dabholkar, G.~Gibbons, J.~A.~Harvey and F.~Ruiz Ruiz,
\jn{\NPh}{ B}{340}{33}{1990};
M.~J.~Duff and K.~S.~Stelle,
\jn{\PhL}{ B}{253}{113}{1991}.



\bibitem{sev-massv}
S. J. Poletti, J. Twamley and D. L. Wiltshire, 
\jn{\PhR}{ D}{51}{5720}{1995};
\jn{\CQG}{}{12}{1753}{1995};
A.~Lukas, B.~A.~Ovrut and D.~Waldram,
\jn{\NPh}{ B}{495}{365}{1997}.


\bibitem{will}
C. Will,
\boo{Theory and experiment in gravitational physics}{Cambridge University 
Press}{Cambridge}{1981}{}


\bibitem{susypot}
S.~J.~Gates and B.~Zwiebach,
Phys.\ Lett.\ B {\bf 123}, 200 (1983);
N.~P.~Warner,
Nucl.\ Phys.\ B {\bf 231}, 250 (1984);
C.~M.~Hull and N.~P.~Warner,
Nucl.\ Phys.\ B {\bf 253}, 675 (1985);
L.~Castellani, A.~Ceresole, R.~D'Auria, S.~Ferrara, P.~Fre and E.~Maina,
Phys.\ Lett.\ B {\bf 161}, 91 (1985);
K.~T.~Mahanthappa and G.~M.~Staebler,
Z.\ Phys.\ C {\bf 33}, 537 (1987);
R.~D'Auria, S.~Ferrara and P.~Fre,
Nucl.\ Phys.\ B {\bf 359}, 705 (1991).





\end{references}
\end{document}